\documentclass[]{aa}

\usepackage{graphicx}
\usepackage{txfonts}
\usepackage{lipsum}
\usepackage{subcaption}         
\usepackage{lscape}             
\usepackage{placeins}           

\usepackage[breaklinks=true,colorlinks=true,linkcolor=blue,urlcolor=blue,citecolor=blue]{hyperref}

\newcommand{\msun}{{\rm M}_{\odot}}

\newcommand{\yr}{{\rm yr}}

\newcommand{\myr}{{\rm Myr}}

\newcommand{\pmm}{PM\@\xspace}
\newcommand{\es}{ES\@\xspace}
\newcommand{\esoc}{ES$_\mathrm{C}$\@\xspace}
\newcommand{\eg}{e.g.\@\xspace}

\newcommand{\ie}{i.e.\@\xspace}

\begin{document}
\title{1D stellar mergers: entropy sorting and PyMMAMS}
\author{Max Heller\inst{\ref{HITS},\,\ref{UNIHD}}\thanks{maxheller@mailbox.org}, Fabian R.~N.~Schneider\inst{\ref{HITS},\,\ref{ZAHARI}}, Jan Henneco\inst{\ref{NCL},\,\ref{HITS}}, Vincent A.~Bronner\inst{\ref{HITS},\,\ref{UNIHD}} \and Mike Y.~M.~Lau\inst{\ref{HITS},\,\ref{UNIHD}}}
\authorrunning{M.~Heller\inst{\ref{HITS},\,\ref{UNIHD}}, F.~R.~N.~Schneider\inst{\ref{HITS},\,\ref{ZAHARI}}, J.~Henneco\inst{\ref{NCL},\,\ref{HITS}}, V.~A.~Bronner\inst{\ref{HITS},\,\ref{UNIHD}} \and M.~Y.~M.~Lau\inst{\ref{HITS},\,\ref{UNIHD}}}
\institute{
            Heidelberg Institute for Theoretical Studies, Schloss-Wolfsbrunnenweg 35, 69118 Heidelberg, Germany \label{HITS}
            \and Universität Heidelberg, Department of Physics and Astronomy, Im Neuenheimer Feld 226, 69120 Heidelberg, Germany \label{UNIHD}
            \and Zentrum für Astronomie, Astronomisches Rechen-Institut (ZAH/ARI), Heidelberg University, Mönchhofstr.~12-14, 69120 Heidelberg, Germany \label{ZAHARI}
            \and School of Mathematics, Statistics and Physics, Newcastle University, Newcastle upon Tyne, NE1 7RU, United Kingdom\label{NCL}
}
\date{Received xxxx / Accepted yyyy}
\abstract{Stellar multiple systems are the norm, not the exception, with many systems undergoing interaction phases during their lifetimes. A subset of these interactions can lead to stellar mergers, where the two components of a stellar binary system come close enough to coalesce into a single star. Accurately modeling stellar mergers requires computationally expensive 3D methods, which are not suited for exploring large parameter spaces as required \eg, by population synthesis studies. In this work, we compare two 1D prescriptions based on the concept of entropy sorting to their 3D counterparts. We employ a basic entropy sorting method ('\es'), which builds the merger remnant by sorting the progenitor stars' shells by increasing entropy, and a Python version of the 'Make Me A Massive Star' code ('\pmm'), which additionally applies a shock-heating prescription calibrated on SPH simulations of stellar head-on collisions. Comparing to a set of 39 more recent SPH head-on collisions different from the ones used for \pmm calibration, we find that \pmm reproduces the outcome of these mergers a lot better than \es in terms of thermal and composition structure post-merger.

Both 1D methods produce remnants that are rejuvenated more strongly than expected for massive stars, indicating that increased amounts of hydrogen are being mixed into the core. In an effort to further improve \pmm, we introduce a scaling factor for the shock-heating. We compare 1D models with both down- and upscaled heating to a 3D magnetohydrodynamic (MHD) $9 + 8\,\msun$ merger of main-sequence stars. Decreasing the shock-heating improves the agreement in terms of the entropy profile, but has only a minor impact on the subsequent stellar evolution of the remnant. We find that 1D methods are able to approximate 3D stellar merger simulations well, and that shock-heating has to be considered to properly predict the post-merger structures.}
\keywords{Methods: numerical -- Stars: binaries, massive, evolution}
\maketitle
\nolinenumbers
\section{Introduction}\label{sec_introduction}
Most massive stars reside in binary and higher-order systems \citep{Moe2017,Offner2023}, a significant fraction of which are expected to undergo a stellar merger event during their lifetimes \citep{Henneco2024a}. The resulting merger products can \eg appear as stars with ages younger than stars in the same cluster \citep{Schneider2016}, or as blue supergiants (BSGs), spending a significant amount of time in the Hertzsprung Gap (HG) area of a genuine single star with a similar mass \citep{Justham2014,Menon2024,Schneider2024}. Recent studies propose that the stars created in stellar mergers may be detectable in asteroseismic observations \citep{Henneco2025}. Understanding the structure of stellar merger products is therefore crucial for interpreting observations of massive stars, as these can significantly affect the further evolution and therefore observational signatures of these objects.

Simulations of stellar mergers are often performed with 3D methods such as SPH \citep{Glebbeek2013} and 3D MHD methods \citep{Schneider2019}. These approaches, however, are computationally expensive, and therefore not applicable to large populations of merger models. Furthermore, certain types of mergers, such as systems containing an expanded post-MS star, are inaccessible to 3D methods due to the significant difference in scale between primary and secondary progenitor star and the need for prohibitively small timesteps due to the high densities encountered in the post-MS star's core. For these reasons, 1D stellar merger methods have been developed \citep{Podsiadlowski1992,Lombardi2002,Ivanova2002,Gaburov2008,Justham2014,Menon2017,Menon2024,Schneider2024}. These include schemes like entropy sorting (\es) and 'Make Me A Massive Star'. Both methods exploit the principle that in Schwarzschild stable regions, specific entropy $s$ increases monotonically with radius $r$ \citep{Landau1959,Bisnovatyi2002}, $\mathrm{d}s/\mathrm{d}r > 0$ (see also Appendix \ref{sec_entropy_formalism}). 
The first of our 1D merger approximation schemes is 'Make Me A Massive Star' (\texttt{MMAMS}), originally presented in \citep{Gaburov2008}. By comparing the entropic variable $A$, which is related to the specific entropy, between the initial and final state of a range of SPH head-on collisions, they find a shock-heating prescription that can be applied to the progenitor stars before merging them through entropic variable sorting. 
As a competing 1D scheme we also perform mergers using the simple entropy sorting (\es) method, where the progenitor stars' shells are stacked by increasing specific entropy to create the merger remnant. 
In this work, we employ both \es and '\texttt{PyMMAMS}' (\pmm), \texttt{MMAMS} rewritten in Python, for performing stellar mergers. \pmm was created for allowing easier modifiability and community access to the \texttt{MMAMS} stellar merger prescription. We compare the outcome of both 1D merger methods to benchmark simulations of stellar mergers performed with 3D methods to gauge the accuracy of the comparatively simple 1D schemes. First, we consider the SPH simulations of head-on collisions presented in \citet{Glebbeek2013}. The mergers of a $10\,\msun$ primary star with its $1\,\msun$ companion at half-age main sequence (HAMS) and a $20\,\msun$ core-hydrogen-exhaustion (CHEX) with a $8\,\msun$ star of the same age are used to compare the codes in detail. Additionally, the remaining SPH head-on collisions are also recreated with the 1D codes, and we compare which of the progenitor stars sinks to the center of the merger remnant. Next, we compare our merger prescriptions to a 3D MHD simulation of a $9 + 8\,\msun$ main sequence (MS) merger \citep{Schneider2019} using the AREPO code \citep{Springel2010}. We introduce a modification factor into \pmm's shock-heating procedure and study the effects of increasing and decreasing the amount of heating on the post-merger structure and evolution.

The structure of this paper is as follows. In Sect.~\ref{sec_methods} we present the two 1D methods and the MESA setup used for generating the progenitors as well as evolving the remnant stars. Sect.~\ref{sec_results} compares core-ownership, merger product structure and evolution between the 1D methods and the benchmark SPH and 3D MHD simulations. Finally, in Sects.~\ref{sec_discussion} and \ref{sec_conclusion} we discuss the results and summarize our conclusions.

\section{Methods}\label{sec_methods}
\subsection{Entropy sorting}\label{sec_entropy_sorting}
Our most basic 1D stellar merger method works by applying the prescription described in Sect.~\ref{sec_entropy_formalism} to the structures of the merger progenitor stars. Entropy sorting (abbreviated to '\es' in the following) builds the merger product by stacking the progenitor stars' shells by increasing entropy. Mass loss is applied by removing the corresponding amount of matter from the remnant's surface. Since shells of similar entropy but vastly different composition can end up next to each other in the merger product, a remeshing algorithm has to be applied to the raw merger outcome to smooth out large shell-to-shell variabilities in the composition profile. We have created a remeshing scheme, which is outlined in more detail in Sect.~\ref{sec_pymmams}.

The entropy profiles of the progenitors and the resulting merger remnant are not modified during the merger, which can lead to unphysical values of the central density and temperature of the merger remnant after relaxation in MESA (Sect.~\ref{sec_headon}). The specific entropy of an ideal gas is given by the Sackur-Tetrode equation \citep{Sackur1911, Tetrode1912a, Tetrode1912b, Sackur1913},
\begin{equation}
    s = \frac{\mathcal{R}}{\mu} \ln\left(\frac{T^{3/2}}{\rho}\right) + \mathrm{const.}
\end{equation}
with the mean molecular weight $\mu$, temperature $T$ and density $\rho$. Assuming thermal equilibrium allows for employing the scaling relationship between a star's total mass $M$, central temperature and central density (subscript 'c'),
\begin{equation}
    T_\mathrm{c} \propto M^{2/3}\rho_\mathrm{c}^{1/3},
\end{equation}
allows us to find the following relationship between a star's central entropy $S_\mathrm{c}$, its mass $M$ and central temperature $T_\mathrm{c}$:
\begin{equation}
    S_\mathrm{c} \propto \frac{1}{\mu_\mathrm{c}} \ln \left(\frac{M^2}{T_\mathrm{c}^{3/2}}\right).
\end{equation}
More massive stars will therefore feature higher central entropies. Progressing stellar evolution raises both the central temperature and mean molecular weight, so further evolved stars will have lower central entropies than younger stars of the same mass. This is important for entropy sorting mergers because the mass of the merger product can be up to double that of the primary progenitor, while the central entropy is that of the progenitor star with the lower central entropy. The merger remnant therefore has an entropy that is too low for its new mass, leading to a compensatory thermal adjustment, which raises the central temperature. We show in Sect.~\ref{sec_headon} that the \es merger remnants have central densities and temperatures much higher than the benchmark models. To remedy this, we also present merger remnants where only the \es composition structure was used in the stellar engineering procedure (see Sect.~\ref{sec_engineering}). We designate such models as '\esoc'.

\subsection{(Python) Make Me A Massive Star}\label{sec_pymmams}
The 'Make Me A Massive Star' code, originally presented in \citet{Gaburov2008}, approximates stellar mergers in 1D by employing buoyancy sorting and a shock-heating prescription, calibrated on SPH stellar merger simulations of systems with varying primary mass, evolutionary stage and mass ratio. It is currently contained in the AMUSE framework \citep{PortegiesZwart2009,PortegiesZwart2013,Pelupessy2013,PortegiesZwart2018,PortegiesZwart2019}. We have rewritten this code in Python for easier access and modifiability. We use this rewritten version, which we refer to by `\pmm', for all of the mergers presented in this paper. For completeness, we quickly summarize how the code performs stellar mergers below:
\begin{enumerate}
    \item The progenitors are loaded from files and their entropic variable profiles computed. The entropic variable, 
    \begin{equation}
    A = \frac{\beta P}{\rho^{5/3}}\exp^{\frac{8}{3}\frac{1-\beta}{\beta}}
    \label{eq_entropic_variable}
    \end{equation}
    depends on the total pressure $P$, the density $\rho$, and the ratio of gas pressure to total pressure, $\beta$. It is proportional to the specific entropy.
    \item \label{enum_shock_heating}Both progenitors are shock-heated, following the prescription
    \begin{equation}
    \log_{10}\left(\frac{A_\mathrm{n,\,f}}{A_\mathrm{n,\,i}}\right) = a' + b \log_{10}\left(\frac{P_\mathrm{n,\,i}}{P_\mathrm{c,\,i}}\right),
    \label{eq_shock_heating}
    \end{equation}
    where $A_\mathrm{n,\,i}$ and $A_\mathrm{n,\,f}$ are the $n$-th shell's entropic variable before and after shock heating, respectively. $P_\mathrm{n,\,i}$ is the shell's pressure before shock heating, and $P_\mathrm{c,\,i}$ the initial central pressure of the progenitor under consideration. The parameter $a'$ is determined from
    \begin{equation}
    a' = a + \log_{10}(f_\mathrm{heat}).
    \end{equation}
    Both $a$ and $b$ were obtained by \citet{Gaburov2008} by fitting Eq.~\ref{eq_shock_heating} to SPH simulations of stellar head-on collisions involving binaries of different initial masses, mass ratios, and evolutionary stages. $f_\mathrm{heat}$ is a free parameter which is used for adjusting the shock-heating to ensure energy conservation (see step \ref{enum_energy_conservation}). Finally, the progenitor stars' shells are sorted by increasing entropic variable $A$.
    \item \label{enum_ode_solver_he}The merger remnant is built by employing a shooting method as described in \citet[Chapter 12.1]{Kippenhahn2012}, starting in the star's center. A first estimate for the central pressure of the merger product is obtained by averaging the central pressures of the progenitor  stars. This serves as an inner boundary condition for an ODE solver that tries to solve the equation for hydrostatic equilibrium for the merger remnant, with the outer boundary condition being vanishing pressure at the star's surface. Starting at the remnant's center, the densities of the innermost progenitor shells at the new pressure are computed. The denser shell is chosen as the first shell of the merger product\footnote{Since the density of the progenitor shell at the new pressure in the merger remnant is the determining factor of which progenitor shell is chosen, inversions in the entropic variable profile are possible. The density profile, however, will be monotonically decreasing with increasing radius.}. This is repeated until the merger product reaches the combined mass of the progenitor stars minus the mass loss. The surface pressure of the merger product is then used to adjust the estimate for the central pressure, \ie a positive surface pressure means that the central pressure is reduced and vice versa. Only when the surface pressure vanishes (within $0.1$\%), the code moves on to the next step.
    \item \label{enum_energy_conservation}The total energy (gravitational + thermal) of the progenitor stars, $E_1$ and $E_2$, is compared to that of the merger product, $E_\mathrm{rmn}$, and the kinetic energy used to carry away the dynamical ejecta, $E_\mathrm{ej}$. Ideally, the difference in energy before and after the merger vanishes:
    \begin{equation}
        (E_1 + E_2) - (E_\mathrm{rmn} + E_\mathrm{ej}) = 0.
    \end{equation}
    For the ejected matter, $M_\mathrm{ej}$, escape velocity $v_\mathrm{esc}$ from a remnant of mass $M_\mathrm{rmn}$ and radius $R_\mathrm{rmn}$ is assumed:
    \begin{equation}
        E_\mathrm{ej} = \frac{1}{2} M_\mathrm{ej} v_\mathrm{esc}^2,\ \mathrm{with}\ v_\mathrm{esc} = \sqrt{\frac{2GM_\mathrm{rmn}}{R_\mathrm{rmn}}}
    \end{equation}
    Because the code assumes a perfect head-on collision of two stars on a parabolic orbit, the orbital energy vanishes and therefore does not enter the energy considerations above. The shock heating is then adjusted through the parameter $f_\mathrm{heat}$ and the code begins again at step \ref{enum_shock_heating}. Once the relative deviation between initial and final total energy is less than $0.1$\%, the code moves on to the final step.
    \item \label{enum_mixing}The merger product's final structure is computed following step \ref{enum_ode_solver_he}, but now at much higher resolution. Especially in mergers of very differently evolved stars, mass shells of vastly different composition can end up next to each other in the merger remnant. This gives composition profiles that look almost double-valued and will not be easy to import into a stellar evolution code (see Sect.~\ref{sec_engineering}). The merger product's shells are therefore mixed, \ie multiple adjacent shells are combined into one. This alleviates the double-valuedness. Finally, the finished merger product is output to file. 
\end{enumerate}
Our Python version of the code also contains some additions and improvements:
\begin{itemize}
    \item The original mixing procedure described in step \ref{enum_mixing} starts from the center and combines shells until their total mass reaches a predefined threshold. This can lead to situations where shells from the outer region of a pure helium core are combined with hydrogen-rich shells of the adjacent hydrogen-rich envelope, effectively mixing hydrogen into the core and softening the steep composition gradient between core and envelope. Changes in the core mass can have an impact on the stellar evolution of the remnant. We therefore implement a 'remeshing' scheme, which first separates the merger product's structure at steep composition gradients and then applies mixing separately to each region. Finally, the regions are attached again, conserving the composition gradients and not introducing \eg material from the envelope into the core.
    \item Because we are working with MESA models, we added the ability of directly loading MESA profiles and outputting files for stellar engineering (see Sect.~\ref{sec_engineering}) in the stellar evolution code.
    \item The chemical species tracked during the merger can now be specified in the input parameter file. 
    \item We introduce the ability of weakening or strengthening shock-heating (see Sect.~\ref{sec_slow_mergers} for a detailed explanation).
    \item The original code uses a hard-coded prescription for the dynamical mass loss during the merger, also obtained from the SPH head-on collisions used to calibrate the shock-heating. However, the fractional mass loss during head-on collisions is much higher than for slow inspiral mergers of comparable progenitor stars \citep{Schneider2019}. Furthermore, the evolutionary stage of the progenitors and their mass ratio also play a crucial role in how much mass is ejected during the merger \citep{Glebbeek2013}. We therefore add an input parameter that allows for a custom mass loss fraction.
\end{itemize}

\subsection{Stellar engineering and evolution}\label{sec_engineering}
We use MESA \citep{MESA1, MESA2, MESA3, MESA4, MESA5, MESA6} version \texttt{r12778} for simulating the stellar evolution of the merger progenitors and remnants. Importing the merger models into the stellar evolution code is done using MESA's relaxation scheme, which, starting at a ZAMS model, gradually modifies first the composition and then the entropy profile until the differences to the input model fall below a predefined threshold \citep{MESA4}. Since MESA always evolves the stellar model for one timestep after the stellar engineering procedure has been completed, we explicitly disable composition changes from mixing and nuclear burning by setting \texttt{mix\_factor} and \texttt{dxdt\_nuc\_factor} to $0$. This prevents transient, short-lived convective zones from introducing steps in the composition profile. Some models run into timestep-issues during entropy relaxation; we mend such cases by increasing the relaxation timescale \texttt{timescale\_for\_relax\_entropy} and maximum timestep \texttt{max\_dt\_for\_relax\_entropy} from their default value of $10^{-9}\,\yr$, making sure to stay within the order of magnitude of $1$ second. The number of timesteps for the composition relaxation, \texttt{num\_steps\_to\_relax\_composition}, is set to 200. In case of crashes during the engineering procedure (usually caused by issues in \texttt{do1\_energy\_eqn}), this value is raised until the crashes resolve. The number of entropy relaxation timesteps, \texttt{max\_steps\_to\_relax\_entropy}, is kept at $1000$. This ensures that the entropy profile of the input model can be properly replicated. If the engineering procedure reaches this number, the engineered merger product's entropy profile could still be different from the input data. 

For the modeling of the merger remnant's subsequent stellar evolution we employ settings similar to \citet{Henneco2024a} in the case of remnants with masses below $20\,\msun$, and settings taken from \citet{Schneider2024} for merger products with masses equal to or larger than $20\,\msun$. For the remnants below $20\,\msun$, regions of (semi-)convection are identified using the Ledoux criterion. Mixing is approximated by the mixing length theory, with the mixing length parameter $\alpha_\mathrm{mlt}$ set to $2.0$. Semiconvection and thermohaline mixing use $\alpha_\mathrm{sc} = 10.0$ and $\alpha_\mathrm{th} = 1.0$, respectively. Core boundary mixing is included through a step overshooting scheme, which extends convective hydrogen burning cores by $0.20$ scale heights $H_\mathrm{P}$ towards the surface. Convective core-helium and later burning stages have exponential overshooting by $0.015\,H_\mathrm{P}$ applied. Additionally, all convective shells are subject to exponential overshooting of $0.005\,H_\mathrm{P}$. We disable \texttt{use\_dedt\_form\_of\_energy\_eqn} because some merger models got stuck at small timesteps directly post-merger. The setting \texttt{include\_composition\_in\_eps\_grav} is enabled to also capture the impact of composition gradients in the computation of the release/absorption of gravitational potential energy. 

The early evolution of some of our post-MS merger products shows problems with numerical mixing introducing hydrogen into pure helium cores. Composition changes from nuclear fusion and mixing are disabled for the first couple $\yr$ post-merger using \texttt{dxdt\_nuc\_factor} and \texttt{mix\_factor}. A more detailed explanation can be found in Appendix~\ref{sec_numerical_mixing}. 

\section{Results}\label{sec_results}
\subsection{Head-on collisions}\label{sec_headon}
\begin{figure*}
    \includegraphics[width=17cm]{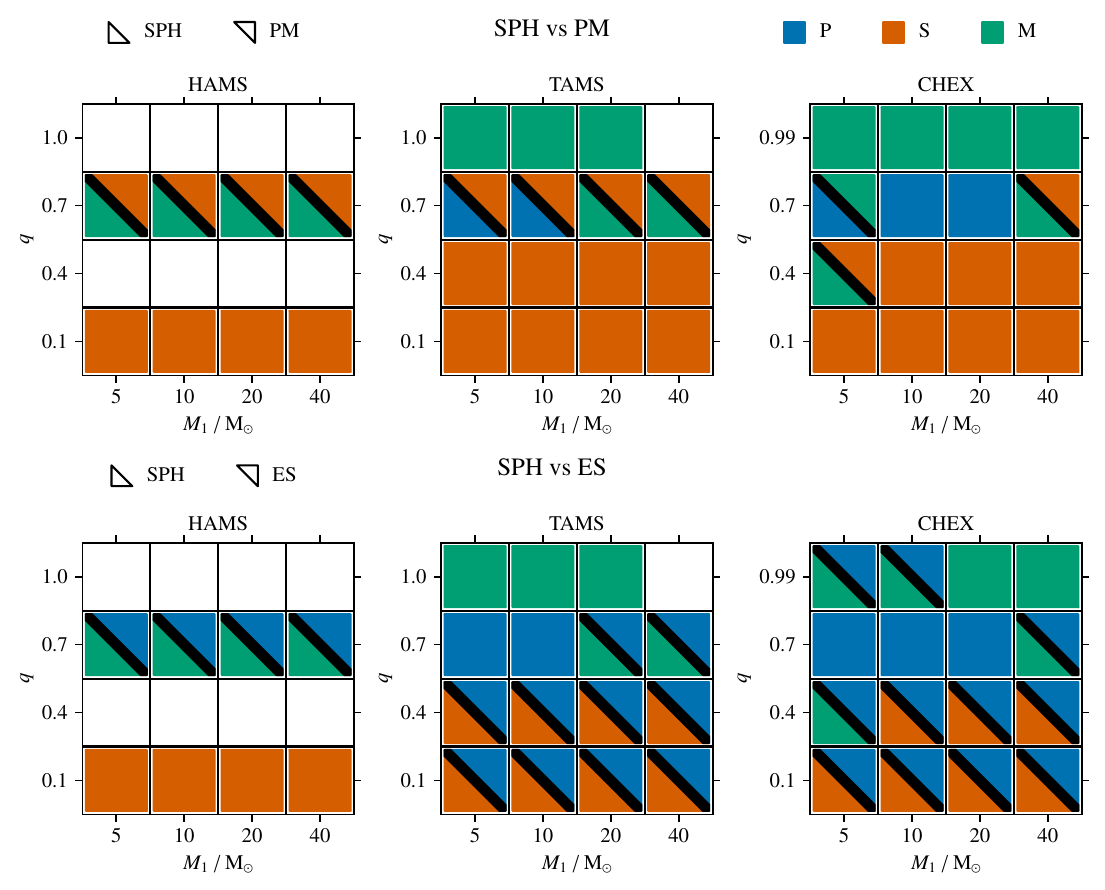}
    \caption{Comparison of the merger case between SPH and \pmm (upper row) and SPH and \es (lower row) for all of the mergers presented in \citet{Glebbeek2013}. The abscissa shows the initial mass of the primary progenitor star, and the ordinate the mass ratio $q = M_2 / M_1$. Each square field corresponds to one merger, and is subdivided into a lower-left (SPH) and upper-right (\pmm/\es) triangle. The color of the triangle indicates the merger case, with blue for case 'P', orange for case 'S', and green for case 'M'. If the merger methods disagree on the case, the corresponding field is additionally marked with a black diagonal bar.}
    \label{fig_case_matrix}
\end{figure*}
We compare the results of our 1D merger remnants obtained from PM and \es to the SPH merger simulations of \citet{Glebbeek2013}. Note that these simulations are different from the SPH simulations used to calibrate the \pmm shock-heating prescription \citep{Gaburov2008} and therefore offer an independent comparison. We evolve our progenitor stars to the same evolutionary stages -- half-age main sequence (HAMS), terminal-age main sequence (TAMS) and core hydrogen exhaustion (CHEX) -- outlined in \citet{Glebbeek2013}. HAMS is defined as being the point in time when half of the physical time between ZAMS and TAMS has passed. TAMS and CHEX are defined by specific points on the HRD, namely the reddest and bluest points of the Henyey hook, respectively. As noted in \citet{Glebbeek2013}, this corresponds to core hydrogen abundances of $\sim 1\%$ for TAMS and $< 0.1\%$ for CHEX. For some of the primary stars we observe smaller central densities at the same evolutionary stage, affecting the merger remnant by changing the entropic variable profile of the progenitor (see Eq.~\ref{sec_entropy_formalism}). The companion stars are evolved to the same physical age as their corresponding primary stars. We do not consider binary interactions prior to the merger, so the progenitor stars represent genuine single stars. We adjust the dynamical mass loss during the merger such that our merger remnants have the same mass post-merger as the SPH head-on collision products. A simple metric for comparing the outcome of the merger simulations is to determine which of the progenitors ends up in the center of the merger remnant. We follow \citet{Glebbeek2013}, who define three distinct merger 'cases' for the outcome: A case 'P' merger remnant has a core predominantly consisting of material coming from the primary progenitor. Conversely, a case 'S' merger remnant has a core mainly made of material coming from the secondary progenitor. Finally, a case 'M' merger remnant has a core that is a mixture of material coming from both progenitors. Using this metric, the merger simulations' outcomes are compared in Fig.~\ref{fig_case_matrix}. We find that \pmm and SPH agree for 28 of 39 merger setups, while \es and SPH agree for only 14 out of the 39 mergers. \es predicts a merger case opposite to the one from the SPH simulations for 15 models, while \pmm does so only for two mergers. Both 1D methods fail at predicting the 'mixed' SPH merger case for the HAMS mergers at a mass ratio $q = 0.7$. This is due to the fact that the progenitors would need to have central entropies / entropic variables (post shock-heating) that are practically identical for the 1D methods to predict the remnant's core to be a mixture of the two progenitor cores. Furthermore, since all of the stars in the HAMS $q = 0.7$ segment have masses $\geq 2\,\msun$, they all feature convective hydrogen burning cores, leading to a flat entropy profile in the central convection region. This further prevents the material from mixing in the 1D simulations, as one of the cores will necessarily have a larger entropy / entropic variable throughout, placing it above the other in terms of mass coordinate post-merger. For the TAMS mergers, \pmm agrees with SPH on the secondary progenitor star ending up in the remnant's center for all of the merger models with mass ratios $q = 0.1$ and $q = 0.4$. This is in stark contrast to the corresponding \es models, which suggest the primary progenitor to occupy the center post-merger. However, for the $q = 0.7$ TAMS mergers \es outperforms \pmm. All of the TAMS $q = 0.7$ \pmm mergers predict a case S, suggesting differences in the primary progenitor morphology at merger as well as deviations in the shock heating prescription being the cause. We observe a similar constellation for the CHEX mergers: For $q = 0.1$ and $0.4$ \es again does not agree with any of the SPH simulations, while \pmm only disagrees for the CHEX 5 + 2 merger. \es manages to correctly predict the outcome of the $q = 0.7$ CHEX mergers with the exception of the most massive progenitor. Finally, \pmm provides the same merger case for all CHEX $q = 0.99$ models as the SPH method, while \es only manages for the $M_1 = 5$ and $10\,\msun$ models. Overall, \pmm deviates least for mergers of systems with $q = 0.1$ and $0.4$. The differences at larger $q$ can be attributed to multiple factors: For large $q$ the progenitor stars will have similar masses and evolve at similar rates, therefore their entropic variable profiles will also be more similar than for low-$q$ systems. This makes it easier for slight miscalibrations in the shock-heating prescription to change the merger case. Because some fine-tuning would be needed to generate central entropic variable values that are similar between the progenitor stars, this effect usually leads to the merger case swinging towards the distinct 'P' or 'S' cases, making it difficult to produce remnants described by case 'M'. \es fails for all but 4 of the systems with $q < 0.5$, where the differences in entropy should be largest due to the much faster evolution of the more massive primary progenitor. Systems with larger mass ratios conversely seem to work better, despite the entropy profiles becoming more similar.

Physically, ES fails at low-$q$ because it preserves the progenitors' entropies: the combined remnant mass is higher but central entropy stays low, forcing the core to over-contract and reorder buoyancy so in some cases the wrong core sinks. \pmm's shock heating corrects this for most low-$q$ systems, but at high-$q$ even small heating offsets matter because the progenitor entropy profiles are similar — tiny changes flip which core is denser at a given remnant central pressure. Core ownership sets the fuel supply and luminosity during core-hydrogen burning, so these mis-assignments propagate into rejuvenation and HRD placement.

It is instructive to compare the composition and thermal structures of the merger remnants obtained from the different merger procedures. Fig.~\ref{fig_hams_10_1} compares the hydrogen mass fraction, temperature, density and specific entropy profiles of the merger remnants computed with the SPH, \pmm and \es merger procedures for a HAMS 10 + 1 system. Note that, due to the plot ranges chosen in \citet{Glebbeek2013}, some of the SPH profiles appear to end at different mass coordinates. For the reasons discussed in Sect.~\ref{sec_entropy_sorting} we compute two kinds of \es models: The basic \es refers to merger models where both the composition and entropy profiles were used during stellar engineering in MESA. The models designated \esoc were obtained by only imposing the composition profile during stellar engineering and letting MESA choose a fitting entropy profile, which will be similar to that of a genuine single ZAMS star of the same mass as the merger remnant \citep{MESA4}. 

\subsubsection{HAMS 10 + 1}
\begin{figure}[htbp]
    \resizebox{\hsize}{!}{\includegraphics{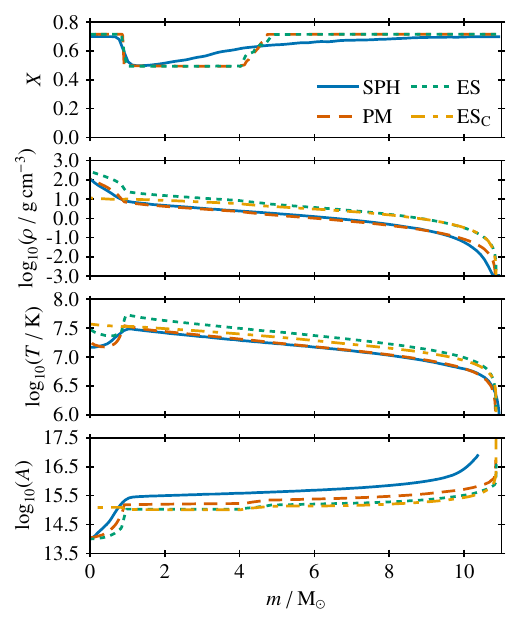}}
    \caption{From top to bottom: Hydrogen mass fraction, density, temperature and entropic variable profiles of the half-age main sequence (HAMS) merger of a $10\,\msun$ primary with a $1\,\msun$ secondary star of the same age. We compare the profiles produced by our 1D methods \pmm, \es and \esoc with the SPH simulations taken from \cite{Glebbeek2013}. The hydrogen mass fraction profile of the \esoc model was not included as it is virtually identical to the \es profile.}
    \label{fig_hams_10_1}
\end{figure}
Fig.~\ref{fig_hams_10_1} compares the merger of an initially $10\,\msun$ half-age main sequence primary star with its initially $1\,\msun$ companion. All three merger methods show similar hydrogen mass fraction profiles, with a roughly $1\,\msun$ core made from hydrogen of primordial abundance. This implies that the central part of the HAMS 10 + 1 is predominantly made from material of the secondary progenitor star, which sinks to the remnant's center due to its low buoyancy. The region above the core shows a hydrogen mass fraction of slightly below $0.5$, meaning that it contains more helium than the merger remnant's core. The increased helium mass fraction is achieved owing to nuclear fusion in the progenitor, which means that the primary progenitor's core ends up above (in mass coordinate) the secondary progenitor in the merger remnant. At the outer edge of this helium-rich region we find the first deviation between the SPH and 1D simulations: While the SPH simulation suggests that the primary's core is distributed throughout a large part of the remnant's envelope, \pmm and \es predict very little mixing with the outer envelope material, as is visible by the primordial hydrogen mass fraction in the outer envelope. 

The density profile shows good agreement between the SPH and \pmm merger products, with almost identical central densities on the order of $10^2\,\mathrm{g\,cm^{-3}}$ and a steeper slope in the core region than in the envelope. The \es merger also displays the different density slopes of the core and envelope, however it predicts densities that are about half an order of magnitude higher than the SPH/\pmm mergers. This is a result of the dependence of the central entropy on the star's mass, as is explained in Sect.~\ref{sec_entropy_sorting}. Because \es does not adjust the entropy profile during the merger, the remnant's (central) entropy is too low for its new mass, and it compensates by increasing its (central) density. This is why we also computed the \esoc merger, where we only impose the composition profile and let MESA choose an appropriate entropy profile. We find that the \esoc merger has a central density that is lower than that of the SPH and \pmm mergers. It is also missing the steep density gradient in the core region, while still overestimating the density in the envelope. 

The temperature profile of the HAMS 10 + 1 merger shows a similar dichotomy as the density profile: The core features a temperature inversion, with the SPH and \pmm models agreeing well in terms of central temperature, and the \es and \esoc mergers predicting significantly higher central temperatures. Again, the \esoc remnant is missing the distinct bipartition into core and envelope.

Finally, the entropic variable profiles of the SPH, \pmm and \es remnants differ only minimally, with a core of very low entropy surrounded by an envelope with only a slightly positive entropy gradient. The \esoc remnant again shows a different structure, with a core of higher entropy than the other remnants, and a more gradual transition between core and envelope. As expected all models show a positive entropy gradient throughout the star, reflecting the basic assumption of convective stability used in all merger methods.

The ES entropy deficit drives 0.5-1 dex higher $\rho_\mathrm{c}$ and hotter $T_\mathrm{c}$ than SPH/\pmm, shrinking the convective core and shifting the star to higher luminosity once H burning resumes. \pmm matches SPH because shock heating lifts the central entropy to a value more appropriate for the remnant's mass. The weak envelope mixing in both 1D methods implies surface abundances will stay near primordial, so any early CNO anomalies seen in SPH would be missed in 1D unless additional mixing is imposed. This can change later in the star's evolution due to dredge-ups.

\subsubsection{CHEX 20 + 8}
\begin{figure}[htbp]
    \resizebox{\hsize}{!}{\includegraphics{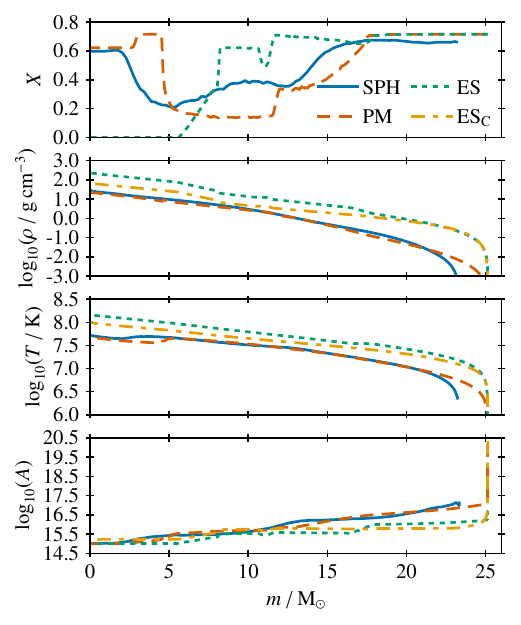}}
    \caption{Same as Fig.~\ref{fig_hams_10_1}, but now for the core-hydrogen exhaustion (CHEX) merger of a $20\,\msun$ primary with an $8\,\msun$ secondary star of the same age.}
    \label{fig_chex_20_8}
\end{figure}
While the merger methods overall agree well for the HAMS 10 + 1 merger, the CHEX 20 + 8 merger shows larger deviations.

This is immediately apparent when looking at the hydrogen mass fraction profile. While SPH and \pmm very roughly agree on the central hydrogen abundance, \es and \esoc predict a pure helium core. From Fig.~\ref{fig_case_matrix} we find that while SPH and \pmm have the secondary progenitor's core end up in the merger product's center, the \es remnant's center is made from material coming exclusively from the primary star's core. This strong difference has a large impact on the further evolution of the merger product. The SPH/\pmm remnant has a hydrogen-rich core similar to a MS star, indicating that it may start core-hydrogen fusion and return to the MS. In contrast, the \es merger's hydrogen mass fraction structure comes closer to that of a post-MS star, and could spend some time as a BSG or even ignite core-helium fusion. 

The density and temperature profiles of SPH and \pmm again agree well, with \pmm only slightly underestimating the density in the inner $10\,\msun$. The deviation of the \es density profile is larger than for the HAMS 10 + 1 merger, now reaching 1 dex. This is caused by the larger mass ratio between primary and secondary progenitor of the CHEX 20 + 8 merger, which means a larger relative change in mass, exacerbating the difference compared to the HAMS 10 + 1 merger.

Comparing the entropic variable profiles of the CHEX 20 + 8 merger shows similar values in the core for SPH, \es and \esoc. Both \es and \esoc show minor entropy inversions, caused by the \es procedure sorting by increasing entropy and not entropic variable, which is missing the contribution of the varying mean molecular weight. SPH and \pmm agree well both in the core and envelope, even giving a similar entropic variable gradient. Both \es methods underestimate the entropic variable in the envelope, also featuring a shallower gradient than the methods including shock-heating.

Here the entropy mismatch has evolutionary consequences: ES places the primary's He core at center, so the remnant behaves like a post-MS star (early He ignition, long BSG phase), whereas SPH/\pmm keep an H-rich core that will lead to a MS phase. The 1 dex central-density excess in ES further accelerates He ignition. This split outcome illustrates that post-MS mergers are highly sensitive to shock heating; using ES alone likely overpredicts BSG incidence and underpredicts rejuvenation.

Comparisons to the remaining detailed mergers presented in \cite{Glebbeek2013} (namely the TAMS 10 + 7, CHEX 20 + 2, CHEX 20 + 14 and CHEX 20 + 19.8 models) can be found in Fig.~\ref{fig_add_mergers}. 

\subsubsection{Rejuvenation of merger remnants}
The violent dynamics of a stellar merger involving main sequence stars can introduce fresh hydrogen into the merger remnant's core, effectively rejuvenating the star and extending its remaining core-hydrogen burning lifetime compared to a genuine single star of the same mass \citep{Schneider2016}. These stars can appear as so-called 'blue stragglers' in the HRDs of stellar clusters, occupying regions beyond the main sequence turn-off point. This effect is also visible in the stellar merger models presented in \citet{Glebbeek2013} and our 1D merger remnants. Similarly to \citet{Glebbeek2008} we compute the relative remaining hydrogen-main-sequence lifetime 
\begin{equation}
    \frac{t_\mathrm{MS}}{\tau_\mathrm{MS}} = 1 - \frac{f_\mathrm{app}}{\alpha},
\end{equation}
and compare it to the apparent fractional age $f_\mathrm{app}$ of the merger remnant. Here, $t_\mathrm{MS}$ is the remaining core-hydrogen burning lifetime of the merger remnant, $\tau_\mathrm{MS}$ the core-hydrogen burning lifetime of a genuine single star of the same mass, $\alpha$ a parameter that quantifies the amount of mixing during the merger and $f_\mathrm{app}$ is defined as
\begin{equation}
    f_\mathrm{app} = \frac{1}{Q_\mathrm{c}(M)} \frac{1}{1 - \phi} \frac{Q_\mathrm{c,\,1} f_1 + Q_\mathrm{c,\,2} f_2 q}{1 + q},
    \label{eq_f_app}
\end{equation}
with $Q_\mathrm{c}(M)$ the fraction of hydrogen consumed during the main sequence evolution of a genuine single star of mass $M$, $\phi$ the fraction of mass lost during the coalescence, and $Q_\mathrm{c,\,1}$, $Q_\mathrm{c,\,2}$, $f_1$ and $f_2$ being the fraction of hydrogen consumed and fractional MS ages of the primary and secondary progenitor stars, respectively. We obtain the $Q_\mathrm{c,\,i}$ and $f_\mathrm{i}$ values from MESA single star models of the respective progenitor masses. 

\begin{figure}[htbp]
    \resizebox{\hsize}{!}{\includegraphics{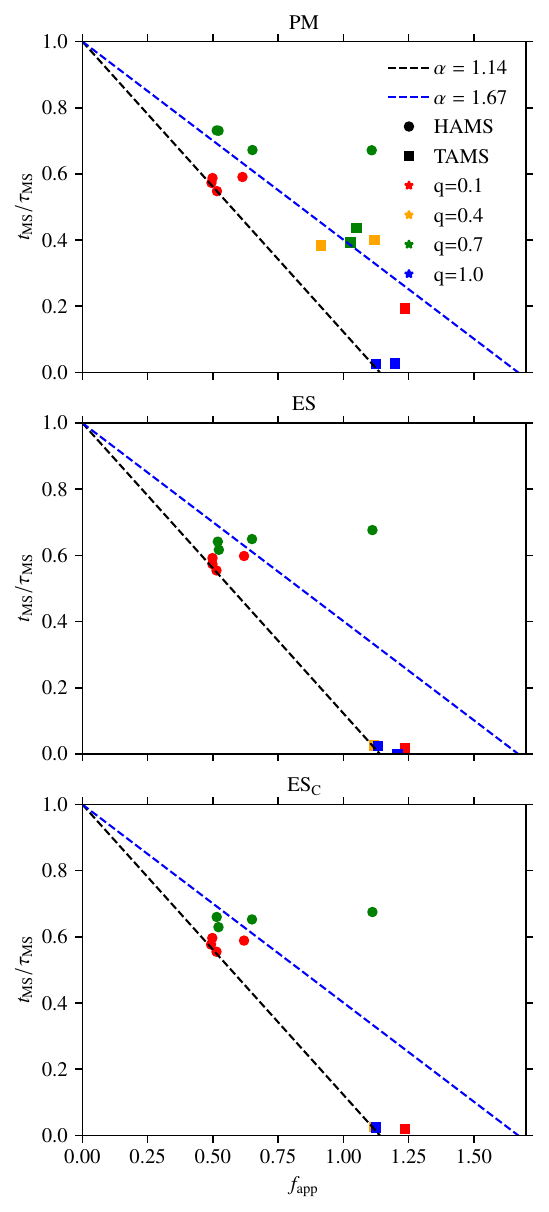}}
    \caption{Relative remaining core-hydrogen burning lifetime $t_\mathrm{MS}/\tau_\mathrm{MS}$ versus apparent fractional age $f_\mathrm{app}$ of the merger remnant for all of our HAMS and TAMS merger models. The black dashed line corresponds to the $\alpha = 1.14$ prescription for low-mass stars offered by \citet{Glebbeek2013}, and the blue dashed line to $\alpha = 1.67$ in \citet{Glebbeek2008} for low-mass stars. HAMS mergers are indicated by circles, TAMS mergers by squares. The different colors correspond to the mass ratio $q$ of the merger. Figure titles again indicate the merger method used.}
    \label{fig_rejuvenation}
\end{figure}

Fig.~\ref{fig_rejuvenation} shows the relative remaining core-hydrogen burning lifetime versus the apparent fractional age of all of our HAMS and TAMS merger models that don't run into numerical issues. The CHEX models are excluded because they are not covered by Eq.~\ref{eq_f_app}. We compare our results to the $\alpha = 1.14$ prescription for low-mass stars offered by \citet{Glebbeek2013} and the $\alpha = 1.67$ prescription from \citet{Glebbeek2008} for low-mass stars. We find that almost all of our merger remnants fall within the range defined by the two $\alpha$ values mentioned above. All of the equal-mass TAMS mergers can be found close to the bottom of the diagram, indicating little amounts of fresh hydrogen being mixed into the core during the merger. This matches our expectation, as the pure helium core of an early post-MS merger product has a lower entropy than the hydrogen-rich core of its companion star. For the HAMS mergers we find a trend of increasing rejuvenation for increasing mass ratio $q$ across all merger methods. This fits the expectation of similar-mass mergers being more violent, allowing for more mixing during the coalescence. There is little difference between the \es and \esoc models in terms of rejuvenation, indicating that the entropy profile does not play a major role in determining the amount of fresh hydrogen mixed into the core during the merger. 

Compared to \citet{Glebbeek2013}, whose merger remnants remaining MS lifetimes scatter around the $\alpha = 1.14$ line, our models follow the $\alpha = 1.67$ line more closely. This implies that our 1D methods overestimate the amount of mixing during the merger compared to the SPH simulations. The \pmm TAMS $q = 0.7$ models, where the secondary progenitor's core occupies the remnant's center, occupy positions close to the $\alpha = 1.67$ line, representing very strong mixing during the merger. As will also be shown in Sect.~\ref{sec_slow_mergers}, the 1D methods show little mixing on smaller scales, \ie larger connected regions of the progenitors are stitched together. However, on larger scales one observes strong mixing, sometimes placing the entire secondary core in the remnant's center, effectively allowing for significant rejuvenation and giving rise to the observed trend. In the \es models the TAMS $q = 0.7$ mergers show less rejuvenation, as the primary progenitor's core ends up in the remnant's center. The trace amounts of hydrogen left are quickly burned off during the early post-merger evolution, leading to little remaining core-hydrogen burning lifetime.

The tendency of our models to cluster near $\alpha = 1.67$ reflects strong mixing of the progenitor material; the $\alpha = 1.14$ seen for the SPH mergers indicates gentler, more local mixing. Both of our 1D methods therefore overestimate mixing during mergers, disproportionately rejuvenating the remnants. 

\subsubsection{Post-merger evolution}
We evolve our merger remnants following the procedure outlined in Appendix~\ref{sec_numerical_mixing} until they reach core helium exhaustion, which we define as the point in time where the central helium mass fraction drops below $10^{-4}$. We did not manage to evolve all of the merger remnants with masses of more than $20\,\msun$ successfully until this point, due to timestep issues either immediately post-merger or later during core-hydrogen burning, despite the use of MESA's MLT++ settings, as outlined in Sect.~\ref{sec_engineering}. We therefore only show the results of the successfully evolved models.

\begin{figure}[htbp]
    \resizebox{\hsize}{!}{\includegraphics{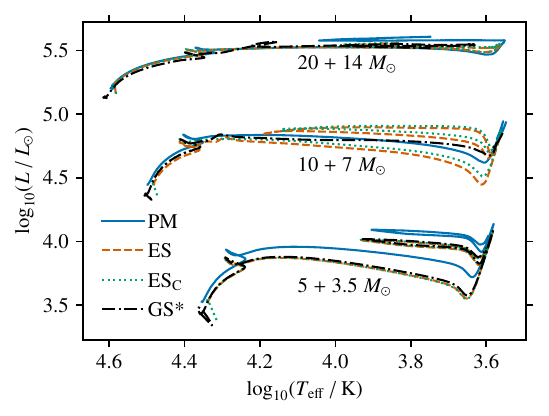}}
    \caption{Hertzsprung-Russell diagram showing the post-merger evolution of the $q = 0.7$ HAMS merger remnants (with the exception of the $M_1 = 40\,\msun$ model) until core helium exhaustion. The track marked with 'GS*' represents a genuine single star with a ZAMS mass equal to that of the merger remnant immediately post-merger.}
    \label{fig_hams_10_7_hrd}
\end{figure}

\begin{figure}[htbp]
    \resizebox{\hsize}{!}{\includegraphics{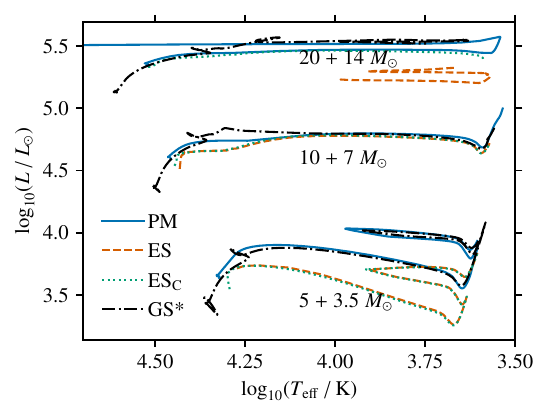}}
    \caption{Same as Fig.~\ref{fig_hams_10_7_hrd} for the corresponding CHEX mergers.}
    \label{fig_chex_10_7_hrd}
\end{figure}

Fig.~\ref{fig_hams_10_7_hrd} shows the post-merger evolution of the $q = 0.7$ HAMS mergers with a $5$, $10$ and $20\,\msun$ primary progenitor in the Hertzsprung-Russell diagram (HRD). The remnants obtained from the different merger methods begin their (post-engineering, post-thermal-relaxation, see Sect.~\ref{sec_numerical_mixing}) evolution at very different points in the HRD (not shown for clarity). This is because our merger methods do not enforce thermal equilibrium (in the case of \es, not even HSE is ensured), and the merger remnants therefore need to thermally adjust to their new mass over a thermal timescale. During this adjustment, all remnants find their way back to the hydrogen-main-sequence, with the different merger methods agreeing well on the effective temperature and luminosity once thermal equilibrium is regained. The subsequent core-hydrogen burning phase happens at similar effective temperatures, but both the \es and \esoc remnants are, to varying degrees, less luminous than the \pmm model for all three primary star masses. This difference in luminosity is enhanced further during the crossing of the Hertzsprung Gap. Comparing to the HAMS 10 + 7 benchmark model in \citet{Glebbeek2013} (their Fig.~7), we find that the point in the HRD at which core-hydrogen burning starts back up is almost identical between the SPH and \pmm methods. Both merger codes give a star that is slightly overluminous when compared to a genuine single star. For the HAMS 5 + 3.5 both \es models have MS tracks that are very similar to the single star. The differences from the GS* tracks also vary with primary mass: While the HAMS 5 + 3.5 \es and \esoc models evolve almost the same as the single star, the HAMS 10 + 7 \es and \esoc remnants are significantly under-luminous compared to the GS* track during the Hertzsprung Gap crossing. For the 20 + 14 merger, \pmm, \es and \esoc agree well on the evolutionary track and are all over-luminous compared to the GS* track.

Differences in shock heating map directly to luminosity: hotter, denser ES cores yield smaller hydrogen burning cores and under-luminous MS/HG tracks at intermediate mass, while PyMMAMS tracks align better with SPH benchmarks. For post-MS mergers, ES's over-dense cores ignite He early, shortening the BSG residence; PyMMAMS retains inert He cores longer, closer to SPH, implying more sustained BSG visibility.

We compare the post-merger evolution of the $q = 0.7$ CHEX mergers in Fig.~\ref{fig_chex_10_7_hrd}. All of the mergers, regardless of method, begin their post-merger evolution as blue supergiants (BSGs) at high effective temperatures. The absence of the characteristic Henyey hook marking the end of the hydrogen-main-sequence implies that the remnants do not return to the main sequence. The \pmm CHEX 5 + 3.5 merger closely follows the GS* track (after the single star has left the main sequence), while both \es models are significantly under-luminous during the HG, RGB and Horizontal Branch phases. The differences between the merger methods become smaller when moving to larger progenitor masses. We can compare our CHEX 10 + 7 models to the TAMS 10 + 7 model in \citet{Glebbeek2013} (their Fig.~10), since both 1D methods predict the same CHEX merger case as the TAMS SPH models (Fig.~\ref{fig_case_matrix}). All three merger methods give an evolutionary track with a slight kink at an effective temperature of around $10^{4.25}\,\mathrm{K}$, most strongly pronounced in the SPH model and least for \pmm. The merger remnants evolve under-luminous through the HG when compared to a genuine single star, and meet up with the GS* track at the bottom of the RGB. 

\begin{figure}[htbp]
    \resizebox{\hsize}{!}{\includegraphics{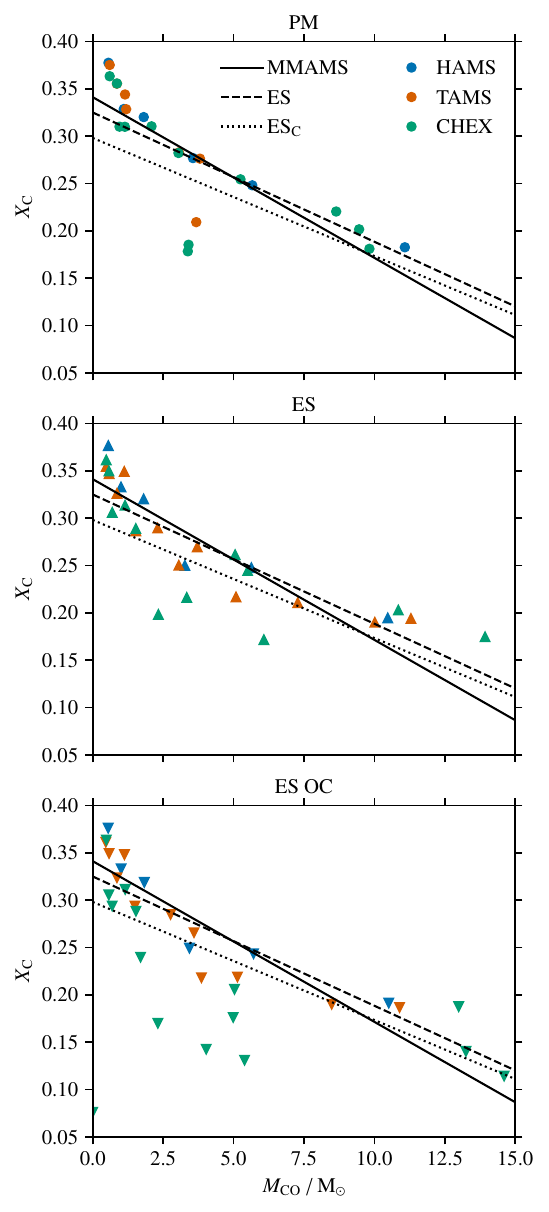}}
    \caption{Central core-carbon mass fraction, $X_\mathrm{C}$, versus carbon-oxygen core mass, $M_\mathrm{CO}$, at core helium exhaustion for all of our merger remnants evolved until this point. The titles of the subfigures indicate the merger method used. We include a linear fit to the data in each panel to offer a qualitative comparison between the 1D prescriptions.}
    \label{fig_carbon_oxygen_core_cheex}
\end{figure}

The merger remnants' core carbon mass fraction, $X_\mathrm{C}$, versus their carbon-oxygen core mass, $M_\mathrm{CO}$, at core helium exhaustion is shown in Fig.~\ref{fig_carbon_oxygen_core_cheex}. We find that regardless of the merger method used, the central carbon mass fraction drops with increasing carbon-oxygen core mass. Both \pmm and \es display some scatter around the general trend, with most outliers having lower $X_\mathrm{C}$ than other remnants of similar core masses. This scatter, however, is not observed for the \esoc models, hinting that the different thermal structures of the merger products are the cause. Furthermore, many of the \es CHEX models have lower $X_\mathrm{C}$ at the same core masses when comparing to \pmm. This is likely caused by the absence of shock heating giving different thermal structures, which is more extreme in mergers involving more evolved progenitor stars because of their lower central entropies. Due to the large scatter, a comparison with the results of \citet{Schneider2021} is difficult. The general appearance of the $X_\mathrm{C}$ versus $M_\mathrm{CO}$ relation is similar, but since core-carbon mass fractions of the different cases (genuine single stars versus stripped stars) are so close together we are hard-pressed to say which of the data is closer to our observations.

\subsubsection{Occurrence of blue supergiants}
As has been shown in previous studies, mergers involving a post-MS star can lead to the formation of long-lived blue supergiants. Our $q = 0.7$ CHEX mergers have the primary's core end up in the remnant's center, giving a core almost exclusively made from helium (Fig.~\ref{fig_case_matrix}). We plot the post-merger evolution of the CHEX 10 + 7 merger in a Kippenhahn diagram in Fig.~\ref{fig_kipp_chex_10_7}. The \pmm remnant starts off with a smaller helium core of only roughly $1\,\msun$, while the \es remnant's core has a mass of around $2\,\msun$. While the core of the \pmm remnant stays radiative and inert until $2\,\myr$ post-merger, the \es remnant ignites core-helium burning almost immediately after the merger. Both remnants feature a convective hydrogen burning shell above the helium core, which fades out about $2.5\,\msun$ after the merger. Shortly after the disappearance of the convective hydrogen burning shell, both remnants develop convective envelopes. We can compare the evolution of our CHEX 10 + 7 models with the evolution of the TAMS 10 + 7 merger presented in \citet{Glebbeek2013} (their Fig.~11), since the change in chemical composition between TAMS and CHEX is negligible, while the slightly different thermal structure of our CHEX progenitors allowed for both \pmm and \es to predict the same merger case as SPH simulations (Fig.~\ref{fig_case_matrix}). The SPH model also features an inert helium core surrounded by a convective hydrogen burning shell. The helium core of the SPH merger also has a mass of around $1\,\msun$, similar to our \pmm model, with the convective hydrogen burning shell extending up to $6\,\msun$ in mass coordinate. The SPH remnant keeps its radiative helium core for more than $3\,\myr$ post-merger, which is longer than both our \pmm and \es models do. It then spends $1.5\,\myr$ core-helium burning, which is slightly shorter than our \pmm model. As a whole, the evolution in the Kippenhahn diagram of our \pmm model agrees well with the SPH model, while the \es remnant ignites core-helium burning too early, while also having a much larger core and shorter lifetime between merger and core-helium exhaustion. 

\subsection{Slow mergers}\label{sec_slow_mergers}
\begin{figure}[htbp]
    \resizebox{\hsize}{!}{\includegraphics{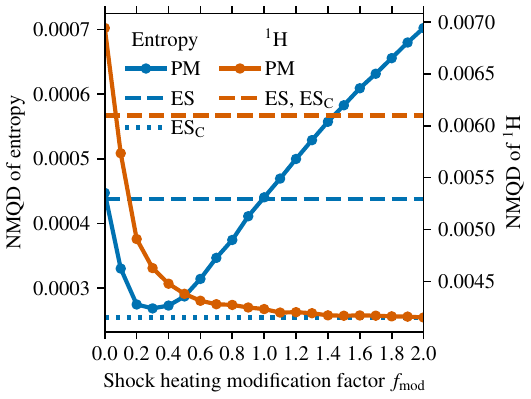}}
    \caption{Normalized mean quadratic deviation (NMQD) of the hydrogen mass fraction and entropy profiles between \pmm models with shock-heating modification factors $f_\mathrm{mod}$ between $0.0$ and $2.0$ and the benchmark 3D MHD merger. We also include the NMQDs of our \es and \esoc models. Because the \es and \esoc models have identical hydrogen mass fraction profiles, only the \es hydrogen NMQD is shown.}
    \label{fig_calibration_nmqd}
\end{figure}
In contrast to the head-on collisions discussed in Sect.~\ref{sec_headon}, stellar mergers in nature are much more likely to follow a slow inspiral. One expects the gradual coalescence of two stars in a slow merger to be less violent than a head-on collision, implying weaker shock-heating and less mass lost dynamically during the interaction. Weakening the shock-heating in \pmm therefore presents a possibility of also simulating inspiral mergers using the 1D prescription. As a benchmark model we use the merger model presented in \citet{Schneider2019}, which was obtained by using the 3D MHD code AREPO \citep{Springel2010}. We use the same progenitor stars and adjust the dynamical mass loss such that it fits the result of the 3D MHD merger. Scaling of the \pmm shock-heating is performed by introducing a modification factor $f_\mathrm{mod}$ into the shock-heating prescription:
\begin{equation}
    a' = a + \log_{10}(f_\mathrm{mod}f_\mathrm{heat}).
\end{equation}
We take special care to completely disable shock-heating for $f_\mathrm{mod} = 0.0$. The modification factor is only applied after the code has determined the final $f_\mathrm{heat}$ (see Sect.~\ref{sec_pymmams}), such that the down- or upscaling is not compensated by the root finding algorithm. We vary $f_\mathrm{mod}$ from $0.0$ to $2.0$ in steps of $0.1$. For every $f_\mathrm{mod}$ we compute the normalized mean quadratic deviation, 
\begin{equation}
    \mathrm{NMQD} = \frac{1}{Q_\mathrm{max} - Q_\mathrm{min}}\sqrt{\frac{\sum_{i=1}^{N}\Delta m_i^2\left(Q_{i,\mathrm{1D}} - Q_{i,\mathrm{3D}}\right)^2}{M_\mathrm{tot}^2}},
    \label{eq_NMQD}
\end{equation}
for both the hydrogen mass fraction and entropy profile. Here, $Q$ is the quantity under consideration, $Q_\mathrm{min}$ and $Q_\mathrm{max}$ are the smallest and largest value for the quantity $Q$ of all the models, respectively, $\Delta m_\mathrm{i}$ is the mass of the i-th shell, and $M_\mathrm{tot}$ is the total mass of the star. The subscripts 1D and 3D refer to the corresponding stellar merger methods. 

\begin{figure}[htbp]
    \resizebox{\hsize}{!}{\includegraphics{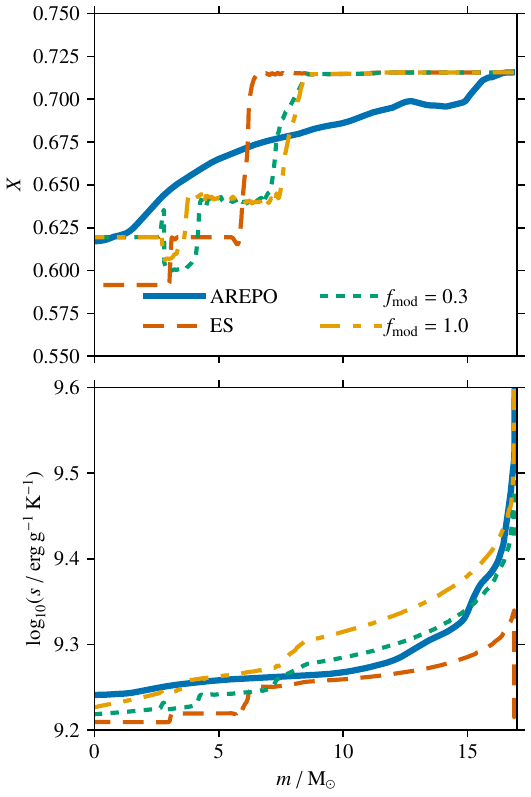}}
    \caption{Top: Comparison of the hydrogen mass fraction profiles between the benchmark AREPO model, the \es model and the \pmm mergers with shock-heating modification factors $f_\mathrm{mod} = 0.3$, $0.5$ and $1.0$. Bottom: Specific entropy profiles for the same models as in the upper plot.}
    \label{fig_calibration_h1_entropy}
\end{figure}

The resulting NMQDs are plotted in Fig.~\ref{fig_calibration_nmqd}, with entropy and hydrogen mass fraction profiles for a selection of differently heated models shown in Fig.~\ref{fig_calibration_h1_entropy}. We find that the hydrogen profile NMQD is largest for the $f_\mathrm{mod} = 0.0$ model, followed by the \es and \esoc mergers. The difference in NMQD between the models despite both not applying any shock heating is caused by the way \pmm performs mergers, which is different from pure entropy sorting (see Sect.~\ref{sec_pymmams}). Further increasing $f_\mathrm{mod}$ leads to an improvement in the NMQD, which however quickly levels out. Increasing the shock heating beyond the basic \pmm prescription (\ie $f_\mathrm{mod} = 1.0$) gives a further slight improvement in the NMQD. This is also visibly when examining the hydrogen mass fraction profiles (upper part of Fig.~\ref{fig_calibration_h1_entropy}). All \pmm $f_\mathrm{mod}$ models have the same central hydrogen mass fraction which only slightly overestimates that of the 3D MHD benchmark merger. The $f_\mathrm{mod} = 0.0$ merger features a region of decreased hydrogen mass fraction just outside the core, at mass coordinates of roughly $3-6\,\msun$. Amplifying the shock-heating shrinks this region, and shifts the step at $6\,\msun$ outwards in mass coordinate. This makes the hydrogen mass fraction profile more similar to that of the benchmark model, explaining the decrease in the NMQD. We also observe that increasing $f_\mathrm{mod}$ beyond $1.0$ only leads to minimal improvements in the NMQD. For all of our 1D merger models we observe a lot less mixing than what is expected from the 3D MHD merger. Since both of our 1D methods are based on entropy/entropic variable sorting one necessarily obtains stellar merger models consisting of stitched-together layers of progenitor star material that are large in mass coordinate relative to the remnant's total mass. Parts of this step-like structure are washed out once core-hydrogen burning restarts and the core turns convective, but it is possible that regions of undermixed material remain in the envelope.

The entropy NMQD becomes smallest for a shock-heating modification factor $f_\mathrm{mod} = 0.3$. Completely disabling shock-heating yields an NMQD that is very close to that of the \es merger, with the small deviation being caused by \pmm using a simplified equation of state. Raising the modification factor leads to an increase in the NMQD, with the original shock-heating deviating by a similar magnitude as the case of no shock-heating. Amplifying the amount of shock-heating beyond $f_\mathrm{mod} = 1.0$ gives ever larger NMQDs, confirming that for matching inspiral mergers one needs to consider weaker heating. The difference between $f_\mathrm{mod} = 0.0$ and $1.0$ becomes apparent in Fig.~\ref{fig_calibration_h1_entropy}, where, despite their similar NMQDs, the $f_\mathrm{mod} = 0.0$ model underestimates the entropy throughout the entire star when compared to the benchmark model, while the $f_\mathrm{mod} = 1.0$ model shows a much larger entropy in the envelope.

The composition NMQD minimum near $f_\mathrm{mod} = 1.0$ but entropy minimum near $\approx 0.3$ reflects the underlying physics: inspirals are gentler than head-on collisions, so less shock heating is expected. Choosing $f_\mathrm{mod} \approx 0.3 - 0.5$ sacrifices a small H-profile mismatch to recover the benchmark entropy stratification, which governs subsequent thermal relaxation and core size. Because layer mixing is still step-like in 1D, inspiral surface-abundance and $\mu$-gradient predictions remain conservative compared to the more mixed AREPO model.

Choosing an appropriate $f_\mathrm{mod}$ is therefore a balance between optimizing the composition and entropy profiles. Generally, one should preferentially try to optimize the composition over the entropy profile, because in any case the star will need to thermally adjust to its new post-merger structure. Furthermore, for mergers of post-MS stars, errors in the \pmm composition profiles could take the form of hydrogen being mixed into an otherwise pure helium core, significantly altering the merger product's further stellar evolution. 

\subsubsection{Evolution until core helium exhaustion}
\begin{figure}[htbp]
    \resizebox{\hsize}{!}{\includegraphics{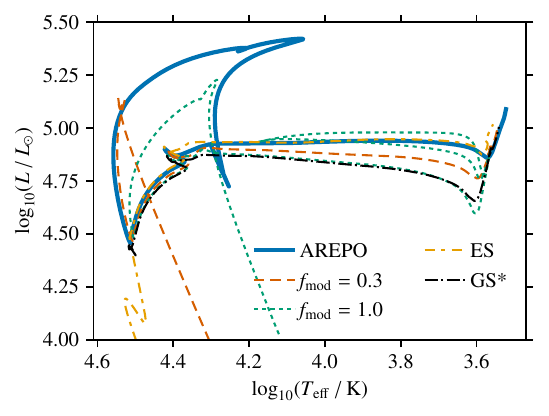}}
    \caption{Evolution of the AREPO, \pmm $f_\mathrm{mod} = 0.3,\,0.5$ and the \es merger remnants. A $16.9\,\msun$ genuine single star (GS*) is also included for comparison.}
    \label{fig_hrd_fmod_arepo}
\end{figure}

Similar to the head-on collision products presented in Sect.~\ref{sec_headon}, we follow the evolution post-merger until core helium exhaustion. Fig.~\ref{fig_hrd_fmod_arepo} shows the evolution of the AREPO benchmark model, the \pmm mergers with shock-heating modification factors $f_\mathrm{mod} = 0.3$ and $0.5$, as well as the \es merger in the HRD until core helium exhaustion. Immediately after the merger event, all models are scattered throughout the HRD due to their different thermal structures, with the $f_\mathrm{mod} = 1.0$ model most closely matching the AREPO merger's path back to the main sequence. Once back in thermal equilibrium and core-hydrogen burning has restarted, all models agree well on the effective temperature and luminosity, effectively overlapping in the HRD. All merger models are slightly over-luminous when compared to a genuine single star of the same mass, with the $f_\mathrm{mod} = 1.0$ model being the least luminous and the \es model the most luminous. In the subsequent Hertzsprung Gap crossing the differences in luminosity are enhanced: The $f_\mathrm{mod} = 1.0$ model is again the least luminous and closest to the track of the genuine single star. Our $f_\mathrm{mod} = 0.3$ fares slightly better, reaching luminosities closer to the AREPO model than the \pmm model with default shock-heating. The \es model is again the most luminous of all merger remnants, and is deviating the least from the AREPO remnant. 

\begin{table}
\caption{Remaining hydrogen-main-sequence lifetime $t_\mathrm{MS}$, helium core mass $M_\mathrm{HE}$, carbon-oxygen core mass $M_\mathrm{CO}$ and central carbon mass fraction $X_\mathrm{C}$ at core helium exhaustion of the 1D merger models obtained with \pmm for different shock-heating modification factors $f_\mathrm{mod}$, as well as the \es and \esoc mergers, compared to the benchmark AREPO model.}
\label{tab_fmod_cheex_properties}
\centering
\begin{tabular}{lcccc}
    \hline\hline
    Model & $t_\mathrm{MS}\,/\,\myr$ & $M_\mathrm{HE}\,/\,\msun$ & $M_\mathrm{CO}\,/\,\msun$ & $X_\mathrm{C}$ \\
    \hline
    AREPO & 9.06 & 6.06 & 4.12 & 0.2078 \\
    $f_\mathrm{mod} = 0.3$ & 8.82 & 5.66 & 3.80 & 0.2588 \\
    $f_\mathrm{mod} = 1.0$ & 8.58 & 5.63 & 3.78 & 0.2441 \\
    ES & 9.33 & 5.90 & 4.01 & 0.2351 \\
    ES$_\mathrm{C}$ & 9.02 & 5.84 & 3.96 & 0.2444 \\
    GS* & 10.84 & 5.60 & 3.74 & 0.2536 \\
    \hline
\end{tabular}
\end{table}

Table~\ref{tab_fmod_cheex_properties} lists the remaining hydrogen-main-sequence lifetime, helium core mass, carbon-oxygen core mass and central carbon mass fraction at core helium exhaustion (which we define as the point at which the central helium mass fraction drops below $10^{-4}$). We find that increasing the shock-heating modification factor $f_\mathrm{mod}$ leads to a decrease in the remaining MS lifetime and helium- and carbon-oxygen core masses. The remaining main sequence lifetime of the benchmark model is most closely matched by the ES$_\mathrm{C}$ merger. Shock-heating causes the merger remnant to have a smaller convective hydrogen core, leading to less fuel being available during core-hydrogen burning. This propagates into smaller helium and carbon-oxygen core masses at core helium exhaustion. All of our 1D merger models overestimate the central carbon mass fraction at core helium exhaustion compared to the AREPO model, implying that the central temperature during core-helium burning is lower than in the benchmark model. 

Raising $f_\mathrm{mod}$ shrinks the convective H core, shortening MS lifetime and reducing He/CO core masses; this is consistent with extra heating puffing up the envelope and lowering central pressure. Even the best-fitting $f_\mathrm{mod}$ cases overestimate $X_\mathrm{C}$ at CHeEX compared to the AREPO merger remnant, signaling cooler helium burning cores in 1D -- an uncertainty that will affect yield and compact-remnant predictions. 

\FloatBarrier
\section{Discussion}\label{sec_discussion}
1D merger methods provide a computationally cheap but accurate approximation for more expensive SPH or 3D MHD simulations of stellar mergers. Furthermore, they offer a way of probing regimes inaccessible to 3D methods, such as post-MS mergers. We have shown that the \pmm code holds up well when comparing to more recent SPH simulations of head-on collisions that are different from the SPH simulations used to calibrate the code's shock-heating prescription. In terms of the merger 'case', \ie which of the progenitor stars occupies the center of the merger remnant, \pmm far outperforms the more simple \es approach, matching the benchmark case in 28 out of the 39 merger setups tested, while \es only does so for 14 setups. Comparing to the SPH models the \es merger remnants also show large deviations in the temperature and density profiles, further highlighting the need for a specialized treatment of entropy changes caused by shock-heating during the merger. Both of the 1D methods, however, tend to produce merger remnants with steep composition gradients and mean molecular weight inversions. Material from the progenitors is often not mixed homogeneously, but placed in the remnant as large chunks, giving step-like structures in the merger product's composition profile. \pmm also does not take into account the composition structure of the progenitors for calculating the entropic variable $A$ during the merger, so (de-)stabilizing effects of composition gradients are disregarded. Still, \pmm delivers better results than the pure \es method, where entropy changes in the stellar material during the merger are not captured at all. \es mergers of systems with similar-mass stars are plagued by a different problem: Since the entropy profile of the merger product is just a composite of the progenitors' entropies for a now much more massive star, the density and temperature are overestimated significantly, giving an unrealistic thermal structure. Only using the composition profile during the post-merger evolution improves this somewhat, as MESA then chooses a thermal profile of a genuine single ZAMS star of similar mass. In this paper, we mainly focused on mergers of massive stars (with masses $\geq 10\,\msun$), similar to those used to calibrate \pmm. Mergers of low-mass progenitor stars, which are more abundant than massive stars and still feature large binary and multiplicity fractions \citep{Offner2023}, might not be described well by the current shock heating prescription, which only considers evolutionary stage and mass ratio for determining which heating parameters to use. 

Introducing a scaling factor in the \pmm shock-heating prescription allows for an improved agreement with 3D MHD simulations of slow inspiral mergers. Weakening the shock-heating leads to better agreement in terms of the entropy structure, while slightly worsening the conformity of the composition profile with the benchmark model. We find that a modification factor $f_\mathrm{mod} = 0.3 - 0.5$ strikes a good balance between optimizing both the composition and entropy profiles with regard to the 3D MHD model. This modification factor, however, is only valid for other high-$q$ MS mergers, and most probably will not hold for other parameter sets, \eg for early post-MS mergers of stars with very different masses. 

Both of our 1D merger methods disregard important physics, such as rotation and magnetic fields. While rotation could be implemented in a simplified manner by assuming angular momentum conservation for every shell during the merger, magnetic fields are much more difficult to include in a 1D framework. Magnetic fields can be generated during the merger process itself \citep{Schneider2019, Ohlmann2025}, but also influence the transport of angular momentum and chemical species in the post-merger evolution. These effects in turn will have an impact on the continued stellar evolution post-merger, how much the star is rejuvenated \citep{Glebbeek2008,Glebbeek2013,Schneider2016} and on the structure at core helium exhaustion (namely the carbon-oxygen core mass and central carbon mass fraction). The structure at CHeEX is important for the explodability, and therefore impacts the final fate of the star \citep{Schneider2021}. For improving the calibration of 1D merger methods, more 3D simulations of MS and early post-MS mergers are needed, especially using more modern approaches including the effects of magnetic fields and ones that capture shocks better than SPH methods. 

\FloatBarrier
\section{Conclusions}\label{sec_conclusion}
We compared merger products obtained from 1D methods with the results of SPH simulations of stellar head-on collisions and the 3D MHD simulation of a slow inspiral merger of MS stars. The 1D code \pmm outperforms the more simple entropy sorting method by a large margin when comparing to the 3D simulations. \pmm correctly predicts which progenitor sinks to the core of the merger remnant in twice as many of our test mergers as \es, while also providing density and temperature profiles that align more closely to the SPH benchmark models than \es. Since \es does not adjust the entropy profile during the merger it predicts much larger central densities and temperatures than the SPH models. The inclusion of a shock heating prescription in \pmm significantly improves upon this shortfall, giving thermal structures much more in line with the 3D results. This is especially important for mergers of stars of very different masses, which is modeled much more accurately by \pmm than \es. Mergers at TAMS and later are also approximated better by \pmm, with \es often wrongly predicting which of the progenitors will occupy the merger remnant's center. 

When comparing to a 3D MHD simulation of a MS $9 + 8\,\msun$ merger, however, \es more closely replicates the post-merger evolution of the remnant, with a more accurate estimate for the helium and carbon-oxygen core masses than \pmm. Weakening the shock-heating in \pmm by introducing a modification factor allows for a better match to the 3D MHD model, with a factor of $f_\mathrm{mod} = 0.3 - 0.5$ providing a good compromise between optimizing the composition and entropy profiles. This implies that 'classical' stellar mergers where the stars inspiral slowly into each other result in less shock-heating than violent head-on collisions, changing the structure and subsequent evolution of the merger remnant.

As more 3D MHD simulations of mergers become available, future studies could probe the applicability of \pmm to mergers of low-mass or early post-MS stars. These can also be used to improve the calibration of the shock-heating prescription, possibly introducing new parameters for different regimes of stellar mergers. Including further physics, such as rotation and magnetic fields, in 1D merger methods also presents an avenue for improving the accuracy of these 1D merger methods.

\begin{acknowledgements}
    We thank Philipp Podsiadlowski for valuable discussions, T.~Neumann, whose bachelor thesis titled 'Probing merger methods in $9\,\msun$ and $8\,\msun$ binary stars' (University of Heidelberg, 2022) laid the foundation for this work, and Utkarsh Basu for supplying the MLT++ inlists. Additional software used includes \texttt{NumPy} \citep{Harris2020}, \texttt{Matplotlib} \citep{Hunter2007}, \texttt{SciPy} \citep{Virtanen2020}, \texttt{PyMesaReader} \citep{Wolf2017} and \texttt{mkipp} \citep{Marchant2017}. JH is grateful for support from UK Research and Innovation (UKRI) in the form of a Frontier Research grant under the UK government's ERC Horizon Europe funding guarantee (SYMPHONY; PI Bowman, grant number: EP/Y031059/1).
\end{acknowledgements}

\section{Data Availability}
Our unified Python framework \texttt{PyStellarMerger}, containing both \pmm and \es, is available on GitHub\footnote{\url{https://github.com/HITS-SET/PyStellarMerger}}. The MESA inlists used for generating the merger progenitors and evolving the merger remnants will be made available on Zenodo. We also provide the structures of all merger remnants as produced by the two 1D methods.

\bibliographystyle{aa}
\bibliography{references}

\clearpage
\begin{appendix}
\section{Entropy in an idealized stellar model}\label{sec_entropy_formalism}
Both \es and PM are based on the observation that in Schwarzschild stable regions of a star, the entropy monotonically increases between core and surface. The equivalence between the Schwarzschild criterion for convective stability and a strictly positive entropy gradient, $\mathrm{d}s/\mathrm{d}r > 0$, can easily be demonstrated by starting from the stability criterion:
\begin{equation}
    \nabla_\mathrm{ad} > \nabla_\mathrm{rad}.
\end{equation}
Expanding the gradients gives
\begin{equation}
    \left(\frac{\partial\ln T}{\partial\ln P}\right)_\mathrm{s} > \frac{\mathrm{d}\ln T}{\mathrm{d}\ln P},
\end{equation}
with $T$ the temperature, $P$ the pressure and $s$ the specific entropy. Multiplying both sides with
\begin{equation}
    -\frac{c_\mathrm{P}}{P} \frac{\mathrm{d}P}{\mathrm{d}r} = \frac{c_\mathrm{P}}{H_\mathrm{P}} > 0
\end{equation}
and expanding the logarithmic derivatives results in
\begin{equation}
    -\frac{c_\mathrm{P}}{T} \left(\frac{\partial T}{\partial P}\right)_\mathrm{s} \frac{\mathrm{d}P}{\mathrm{d}r} > -\frac{c_\mathrm{P}}{T} \frac{\mathrm{d} T}{\mathrm{d} r}
\end{equation}
Using the second law of thermodynamics we can identify 
\begin{equation}
    \frac{c_\mathrm{P}}{T} = \left(\frac{\partial s}{\partial T}\right)_\mathrm{P}.
\end{equation}
Finally, we can apply the reciprocity theorem \citep{Steane2016} to the term on the left-hand side,
\begin{equation}
    -\left(\frac{\partial s}{\partial T}\right)_\mathrm{P} \left(\frac{\partial T}{\partial P}\right)_\mathrm{s} \frac{\mathrm{d}P}{\mathrm{d}r} = \left(\frac{\partial s}{\partial T}\right)_\mathrm{T} \frac{\mathrm{d}P}{\mathrm{d}r},
\end{equation}
and add the term on the right-hand side to both sides of the inequality. This leaves us with
\begin{equation}
    \frac{\mathrm{d}s}{\mathrm{d}r} = \left(\frac{\partial s}{\partial T}\right)_P \frac{\mathrm{d}T}{\mathrm{d}r} + \left(\frac{\partial s}{\partial P}\right)_T \frac{\mathrm{d}P}{\mathrm{d}r} > 0,
    \label{eq_dsdr_expansion}
\end{equation}
which is therefore equivalent to the Schwarzschild criterion for stability.
\section{Anomalous mixing during the thermal relaxation of merger remnants}\label{sec_numerical_mixing}
\begin{figure}[htbp]
    \resizebox{\hsize}{!}{\includegraphics{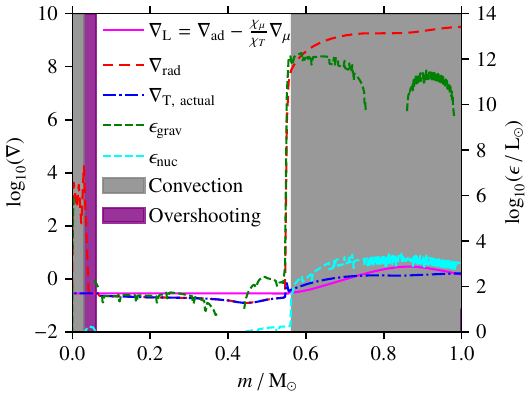}}
    \caption{Ledoux gradient $\nabla_\mathrm{L}$, radiative gradient $\nabla_\mathrm{rad}$ and actual temperature gradient $\nabla_\mathrm{T,\,actual}$ in the inner $1\,\msun$ of the CHEX 10 + 7 remnant roughly $10^{-8}\,\yr$ post-merger. The second ordinate plots gravitational ($\epsilon_\mathrm{grav}$) and nuclear ($\epsilon_\mathrm{nuc}$) energy release in $\mathrm{L_\odot}$. Convection and overshooting are indicated by grey and purple background colors, respectively.}
    \label{fig_rejuv_gradients}
\end{figure}

We disable composition changes from mixing and nuclear fusion during the early evolution of our merger remnants. The reasons for this procedure are illustrated here on the example of a CHEX $10 + 7\,\msun$ merger. The remnant features a pure helium core of about $1\,\msun$ surrounded by a hydrogen-rich envelope. \citet{Glebbeek2013} find this star to keep its inert helium core with a convective hydrogen burning shell for more than $3\,\myr$ post-merger, appearing as a BSG. Despite the \pmm merger product matching the SPH structure well, we observe a very different evolution in the Kippenhahn diagram for the \pmm merger (Fig.~\ref{fig_nomix_rejuv_kipp}). Directly after the merger, there are small convective zones in the center and the inner envelope. The envelope's convective zone immediately starts extending inwards, finally connecting with the central convective zone. This large convective zone connecting core and envelope is maintained until about $10^{-2}\,\yr$ after the merger, and then, starting from the center, dissipates until $1\,\yr$ after the merger. Closer investigation shows that enormous amounts of gravitational potential energy are released at the boundary of core and convective envelope (Fig.~\ref{fig_rejuv_gradients}), raising the local luminosity and therefore the radiative temperature gradient $\nabla_\mathrm{rad}$. The radiative temperature gradient far exceeds the Ledoux gradient, defined as 
\begin{equation}
    \nabla_\mathrm{L} = \nabla_\mathrm{ad} - \frac{\chi_\mu}{\chi_\mathrm{T}} \nabla_\mu,\ \mathrm{with}\ \chi_\mu = \frac{\partial \ln P}{\partial \ln \mu}\ \mathrm{and}\ \chi_\mathrm{T} = \frac{\partial \ln P}{\partial \ln T}.
\end{equation}
As a consequence, the area of large $\epsilon_\mathrm{grav}$ becomes Ledoux-unstable, causing convection. The convective zone extends inwards, eventually connecting core and envelope and mixing hydrogen into the central region. The freshly mixed hydrogen is ignited, further raising the luminosity and stabilizing the convective region. This leads to an erroneous 'rejuvenation' of the post-MS merger product. 

\begin{figure}[htbp]
    \resizebox{\hsize}{!}{\includegraphics{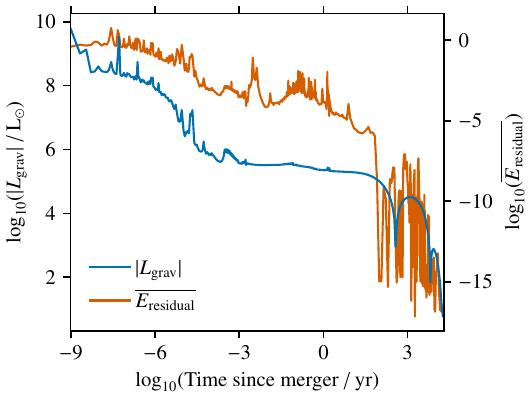}}
    \caption{Absolute gravitational luminosity and energy residual of early thermal relaxation.}
    \label{fig_lgrav_eresid}
\end{figure}

\begin{figure}[htbp]
    \resizebox{\hsize}{!}{\includegraphics{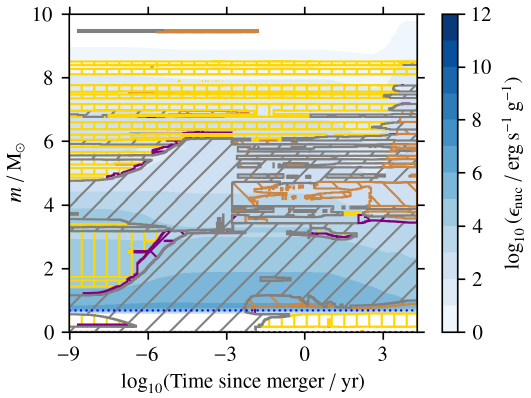}}
    \caption{Kippenhahn diagram of the CHEX 10 + 7 remnant's post-merger evolution where changes in the composition have been disabled.}
    \label{fig_nomix_rejuv_kipp}
\end{figure}

\begin{figure}[htbp]
    \resizebox{\hsize}{!}{\includegraphics{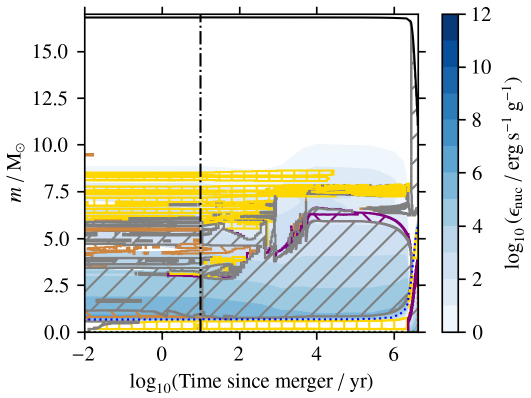}}
    \caption{Kippenhahn diagram of the CHEX 10 + 7 remnant's post-merger evolution. The black dash-dotted line indicates when composition changes are re-enabled.}
    \label{fig_mix_kipp}
\end{figure}

This 'rejuvenation', however, is not the consequence of physical effects, but rather caused by numerical issues raising $\epsilon_\mathrm{grav}$. We therefore disable composition changes until the gravitational energy release subsides and the helium core is convective again (Fig.~\ref{fig_nomix_rejuv_kipp}). The duration of evolution during which composition changes are disallowed needs to be long enough such that no hydrogen can get into the core, and being as short as possible as not to let the merger remnant evolve away from the thermal profile prescribed by the merger method. Tools aiding in this decision are the gravitational luminosity $L_\mathrm{grav}$, which is the integral of $\epsilon_\mathrm{grav}$ over the entire star, and the energy residual $E_\mathrm{residual}$, which is a measure of how well energy is conserved in MESA. Fig.~\ref{fig_lgrav_eresid} shows these metrics for the CHEX 10 + 7 merger discussed above. At about $10^2\,\yr$ post-merger there is a sharp drop in the energy residual. $L_\mathrm{grav}$ becomes smooth already $1\,\yr$ after the merger, indicating a decrease in numerical issues. Restarting the evolution from the model at $10^1\,\yr$ with composition changes enabled yields the evolution presented in Fig.~\ref{fig_mix_kipp}. The envelope convective zone no longer connects to the core, and the star remains in a BSG configuration similar to \citet{Glebbeek2013}.

\section{Additional figures}
\begin{figure*}
    \resizebox{\hsize}{!}{\includegraphics{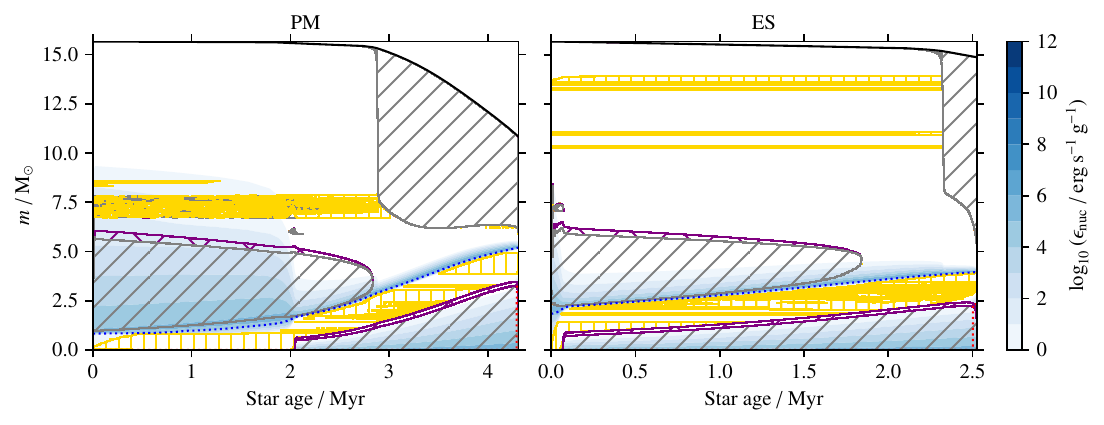}}
    \caption{Kippenhahn diagram of the post-merger evolution of the CHEX 10 + 7 merger remnants obtained with \pmm and \es. Both remnants were evolved until core helium exhaustion. Grey hatching indicates convection, purple and yellow hatching represents overshooting and thermohaline mixing, respectively. The blue background colors refer to nuclear energy production. The blue dotted line describes the extent of the helium core, while the red dotted line signifies the carbon core.}
    \label{fig_kipp_chex_10_7}
\end{figure*}
\begin{figure*}
    \centering
    \resizebox{\hsize}{!}{\includegraphics{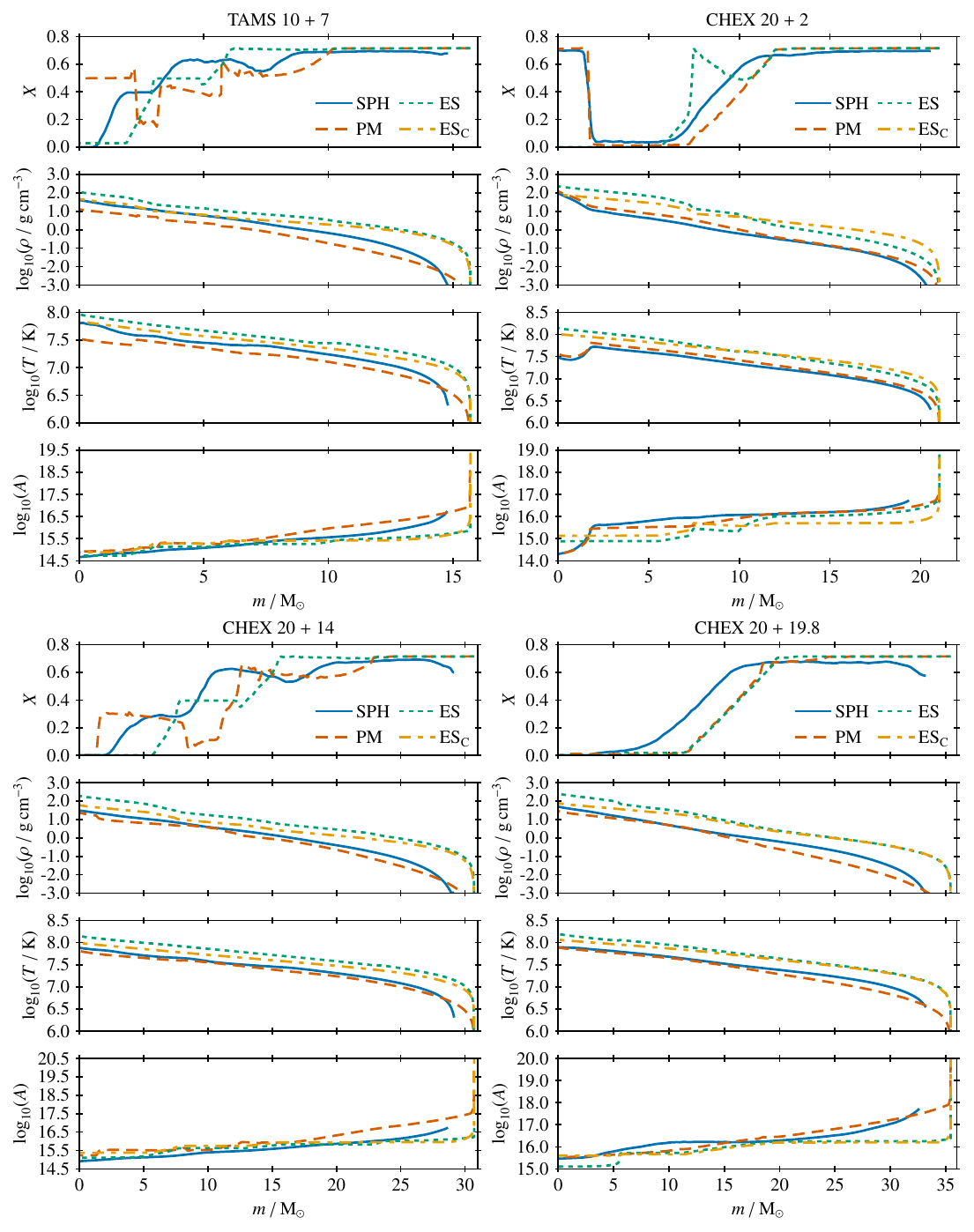}}
    \caption{Additional mergers also present in \citet{Glebbeek2013}.}
    \label{fig_add_mergers}
\end{figure*}
\end{appendix}
\end{document}